\documentclass{svproc}
\usepackage{graphicx}
\usepackage{amsmath}
\usepackage{cite}
\usepackage{tabularx}
\usepackage{authblk}
\usepackage{hyperref}
\hypersetup{
    colorlinks=true,
    linkcolor=blue,
    filecolor=magenta,      
    urlcolor=cyan,
}
\usepackage{url}

\begin{document}
\mainmatter              

\title{Estimating the Sensitivity of IceCube-Gen2 to Cosmic Ray Mass Separation}

\titlerunning{Estimating the sensitivity of IceCube-Gen2 to Cosmic Ray Mass Separation}  

\author[]{Manisha Lohan\\ for the IceCube-Gen2 Collaboration
\footnote{\url{https://authorlist.icecube.wisc.edu/icecube-gen2}}}

\authorrunning{Manisha Lohan}
\institute{Department of High Energy Physics, TIFR, Mumbai-400005, India\\
Email: manisha.lohan@tifr.res.in}

\maketitle             

\begin{abstract}
IceCube-Gen2 is a proposed extension to the existing IceCube Neutrino Observatory at the South Pole. It will consist of three components: an in-ice optical array, a surface array on top of the optical array, and a radio array for detecting ultra-high energy neutrinos. Here we study the sensitivity of this future detector to the mass separation of primary cosmic rays, using CORSIKA Monte Carlo simulations of extensive air showers initiated by H, He, O and Fe primaries. The surface array will use two types of detection technologies: scintillation detectors and radio antennas; the latter are not considered in this study. A set of mass-sensitive variables are investigated utilizing both the scintillators of the surface array and the full optical in-ice array. Among these, the high-energy muons measurable by the in-ice array are found to have the highest mass separation power for showers for which the cosmic-ray energy is known, e.g. from the surface array.

\end{abstract}

\section{Introduction}
Details of origin and acceleration mechanism of high-energy cosmic rays are still largely unknown \cite{cr1, cr2, cr3}. IceCube/IceCube-Gen2 observes cosmic rays indirectly via the detection of air-showers generated in the Earth’s atmosphere \cite{cr4, cr5}. The challenge is to reconstruct from the air-shower measurements the primary cosmic-ray parameters like energy, mass and arrival direction. Various techniques allow us to study the mass composition, e.g. via observation of the electron-to-muon ratio of the secondary particles, shower maximum, or by measuring the muon content of air-showers at different muon energies \cite{cr6,cr7}. IceCube-Gen2 is planned to consist of three components: an optical array with 120 optical strings (between depth of 1344 m to 2689 m), made up of 9600 digital optical modules (DOMs), a surface array on top of the optical array, able to detect cosmic ray air showers up to energies of $\geq$ 10$^{18.5}$ eV, and a radio array sensitive to ultra-high energy neutrinos, having 361 hybrid stations \cite{cr8}. The IceCube-Gen2 surface array will have 130 stations. Each surface array station will have 8 scintillator panels and 3 radio antennas. The horizontal spacing between IceCube-Gen2 strings will be 240 m and vertical spacing between two adjacent DOMs will be 17 m. The geometric aperture of the IceCube-Gen2 surface array for coincident in-ice and surface air-shower events will be $\sim$8 km$^{2}$sr (presently it is 0.25 km$^{2}$sr for IceTop). Scintillators and in-ice array combined will allow to measure particles at the surface and the TeV-muon component in-ice. In the current work, initial cosmic-ray mass separation studies for IceCube-Gen2 are presented studying the combined response of surface array and optical in-ice array.

\section{Monte Carlo datasets}
Mainly four types of thinned \cite{cr9} air-shower datasets, generated by H, He, O and Fe primaries are used for the current studies. These datasets are generated using CORSIKA 7.7400 \cite{cr10} for the same primary energy of 10$^{17}$ eV. These are nearly vertical showers, corresponding to a zenith angle range of, 0$^{\circ}$-19$^{\circ}$. For these simulations, high-energy hadronic interactions are modeled with SIBYLL 2.3d \cite{cr11} and low-energy interactions with FLUKA \cite{cr12}.\\
For the surface array simulations, mainly GEANT4 \cite{cr13} is used and in-ice (optical array) simulations are done using PROPOSAL \cite{cr14}. For the current work, IceCube-Gen2 simulations are done with IceCube optical modules (DOMs) but scaled to the proposed IceCube-Gen2 spacing.\\
The SpiceBFRv1 \cite{ice} ice-model is used that describes the scattering and absorption properties of the ice and how the photons move through the ice.\\
To simulate the trigger for the scintillator array, at-least 5 triggered scintillators are required in order to provide enough sampling points to facilitate good reconstruction and for the optical array, only those DOM hits are considered as good physics hits for which any of their neighbouring DOMs got hits within a time window of 1750 ns.
\section{Analysis method}
Concerning the analysis technique, a Linear Discriminant Analysis (LDA) is performed using scikit-learn package of Python \cite{cr15}. The main goal of LDA is to find the best separation axis between the two data classes. LDA helps to reduce the dimensionality while retaining the information of original variables. The mass separation power is quantified in terms of Figure of Merit (FOM) which is defined as: 
$$\text{FOM}=\frac{ |\mu_p-\mu_i|}{\sqrt{\sigma_{p}^2+\sigma_i^2}}$$
where $\mu$, $\sigma$ are the mean and standard deviation of the mass sensitive parameter; p =  H and i = Fe, He, O etc. 
\section{Mass sensitive observables}
From the scintillator array, the slope ($\beta$) of the lateral distribution function (describing well the scintillator signal) is used as a mass-sensitive parameter \cite{cr16}. The FOM values corresponding to the $\beta$ only case (given in Table \ref{table1}) confirm that it has physical-mass-separation capability. Corresponding to the optical in-ice array, the following variables are studied: charge of in-ice pulses, DOM hits, in-ice muons (estimated using a specific IceCube software named MuonGun \cite{cr17}) and the energy loss of in-ice muons (a proxy variable for in-ice muons). In-ice muons are found to have the highest mass separation power, but reconstructing the exact number of in-ice muons using a IceCube-Gen2 like sparse detector does not seem practically feasible. Therefore, the energy loss of in-ice muons measured by the Cherenkov-light is used as a proxy variable and found to have a correlation with the number of muons \cite{cr18}.

\section{Figure of Merit (FOM) values}
FOM values are calculated corresponding to different combinations of surface and optical variables (measured on event by event basis), shown in Table \ref{table1}. As the IceCube-Gen2 reconstruction resolution for the air-shower observables are not yet fully evaluated, the FOM values calculated in the current work are preliminary, since any kind of systematic uncertainties are not accounted for at this stage.

\begin{table}[h]
\caption{FOM values corresponding to different combinations of variables for Proton vs Iron, Proton vs Oxygen \& Proton vs Helium (work-in-progress).}

\centering
\begin{tabular}{|c|c|c|c|}
\hline

Variables \linebreak & Proton vs Iron \linebreak & Proton vs Oxygen \linebreak & Proton vs Helium\\[0.5ex] 
\hline
$\beta$ & 0.795 & 0.427 & 0.221 \\
\hline
$\beta$, DOM hits & 0.811 & 0.543 & 0.222 \\
\hline
$\beta$, in-ice charge & 0.841 & 0.538 & 0.242 \\
\hline
$\beta$, in-ice muon ($\mu$) (MC truth) & 2.794 & 1.643 & 0.554 \\
\hline
$\beta$, dE (energy loss of in-ice $\mu$) & 1.718 & 0.949 & 0.295\\
\hline
\end{tabular}
\label{table1}
\end{table}

\section{Summary}
The resulting FOM values corresponding to a proton/iron separation for any combination of the surface and optical variables are the highest i.e. showing the maximum separation. A FOM \textgreater 1.5 \cite{cr19} (the confusion between the populations for FOM-1.718 and 2.794 is  $\sim$4\% and $\sim$2\% respectively, indicating a good separation), in particular, confirms that proton and iron can be reasonably separated and is worthwhile to be further studied. A separation of proton and oxygen is more difficult, but also seems feasible. However, the studied variables seem to be insufficient to separate proton from helium primaries. In summary, a combination of IceCube-Gen2 surface and optical variables gives a promising input for cosmic-ray mass separation. At cosmic-ray energies above a few 10 PeV, the radio antennas of the surface array will provide additional mass separation.


\begin{thebibliography}{6}

\bibitem {cr1}
K. H. Kampert et al., Measurements of the cosmic ray composition with air shower
experiments, \textit{Astroparticle Physics 35 (2012) 660-678}. 

\bibitem{cr2}
A. Coleman et al., Ultra high energy cosmic rays: The intersection of the Cosmic
and Energy Frontiers, \textit{Astroparticle Physics 149 (2023) 102819}.

\bibitem{cr3} R. Engel, Indirect Detection of Cosmic Rays. In: Grupen, C., Buvat, I. (eds)
Handbook of Particle Detection and Imaging. Springer, Berlin, Heidelberg, 2012.


\bibitem{cr4} A. Achterberg et al., First Year Performance of The IceCube Neutrino Telescope,
\textit{Astropart. Phys. 26 (2006) 155-173}.

\bibitem{cr5} M. G. Aartsen et al., IceCube-Gen2: A Vision for the Future of Neutrino Astronomy
in Antarctica, \url{https:// arxiv.org/ abs/ 1412.5106}.

\bibitem{cr6} Ewa M. Holt et al., Enhancing the cosmic-ray mass sensitivity of air-shower arrays
by combining radio and muon detectors, \textit{Eur. Phys. J. C (2019) 79:371}.

\bibitem{cr7} Benjamin Flaggs et al., Studying the mass sensitivity of air-shower observables using
simulated cosmic rays, \textit{Phys. Rev. D 109, 042002}.

\bibitem{cr8} IceCube-Gen2 Collaboration et al., IceCube-Gen2 Technical Design Report, \url{https://icecube-gen2.wisc.edu/science/publications/tdr/}.


\bibitem{cr9} A. M. Hillas, Shower Simulation: Lessons from MOCCA, \textit{Nuclear Physics B (Proc.
Suppl.) 52B (1997) 2942}.


\bibitem{cr10} D. Heck et al., Extensive Air Shower Simulation with CORSIKA: A User’s Guide,
\url{https:// www.iap.kit.edu/ corsika/ 70.php}.

\bibitem{cr11} F. Riehn et al., Hadronic interaction model SIBYLL 2.3d and extensive air showers,
\textit{PHYSICAL REVIEW D 102, 063002 (2020)}.

\bibitem{cr12} T.T. Bohlen et al., The FLUKA Code: Developments and Challenges for High Energy and Medical Applications, \textit{Nuclear Data Sheets 120 (2014) 211-214}.


\bibitem{cr13} Geant4 Collaboration et al., Geant4 - A Simulation Toolkit, \url{https://geant4-userdoc.web.cern.ch/UsersGuides/ForApplicationDeveloper/BackupVersions/V10.4/fo/BookForAppliDev.pdf}.

\bibitem{cr14} J. H. Koehne et al., PROPOSAL: A tool for propagation of charged leptons, 
\textit{Computer Physics Communications 184 (2013) 2070–2090}.

\bibitem{ice} The IceCube Collaboration, Light diffusion in birefringent polycrystals and the IceCube ice anisotropy, \textit{PoS ICRC2019 854}.

\bibitem{cr15} \url{https://scikit-learn.org/stable/supervised_learning.html}.

\bibitem{cr16} A. S. Leszczyńska, Potential of the IceTop Enhancement with a Scintillation De-
tector Array, PhD thesis 2020, \url{https://publikationen.bibliothek.kit.edu/1000131245/109171043}.

\bibitem{cr17} Jakob van Santen, Neutrino Interactions in IceCube above 1 TeV, PhD Thesis 2014, \url{https://inspirehep.net/literature/1339582}.

\bibitem{cr18} Tom Feusels, Measurement of cosmic ray composition and energy spectrum between 1PeV and 1EeV with IceTop and IceCube, PhD Thesis 2013, \url{https://biblio.ugent.be/publication/4337238}.

\bibitem{cr19} The Pierre Auger Collaboration et al., The Pierre Auger Observatory Upgrade - Preliminary Design Report, \url{https://doi.org/10.48550/arXiv.1604.03637}.
\end{thebibliography}
\end{document}